\documentclass[a4paper]{aa}
\usepackage{natbib}
\bibpunct{(}{)}{;}{a}{}{,} 
\usepackage{graphicx}
\begin{document}

\title{Parameters of Warm Molecular Clouds from Methyl Acetylene Observations}

\author{A. V. Alakoz\inst{1}    \and
        S. V. Kalenskii\inst{1} \and
	V. G. Promislov\inst{1} \and
        L.~E.~B.~Johansson\inst{2} \and
	A.~Winnberg\inst{2}}

\offprints{A.V. Alakoz, \email{rett@tanatos.asc.rssi.ru}}

\institute{Astro Space Center, Lebedev Physical Instutute, ul. Profsoyuznaya
84/32, Moscow, 117997. Russia \and
Onsala Space Observatory, S-439 92, Onsala, Sweden}

\date{Received / Accepted } 
		      
\abstract{
The results of a survey of 63 galactic star-forming regions in the $6_K-5_K$
and $5_K-4_K$ methyl acetylene lines at 102 and 85~GHz, respectively, are
presented. Fourty-three sources were detected at 102~GHz, and twenty-five at
85~GHz. Emission was detected towards molecular clouds with  kinetic
temperatures 20--60~K (so-called ``warm clouds''). The CH$_3$CCH abundances
in these clouds appeared to be about several units~$\times$~10$^{-9}$.
Five mapped sources were analyzed using the maximum entropy method. The sizes of
the mapped clouds fall within the range between 0.1 and 1.7~pc, virial
masses --- between 90~---~6200~M$_\odot$, and densities --- between 
$6\times 10^4$ and $6\times 10^5$ cm$^{-3}$. The CH$_3$CCH sources spatially
coincide with the CO and CS sources.
Chemical evolution simulations showed that the typical methyl acetylene
abundance in the observed clouds corresponds to an age of
$\approx 6\times 10^4$~years.}

\maketitle
\section{INTRODUCTION}

Massive stars are known to form in so-called "warm" molecular clouds
\citep{Olmi,kalensky1},
which have
temperatures 30~--~50 K, masses from hundreds to thousands $M_\odot$,
and sizes 0.1~--~3 pc. Therefore the exploration of such clouds
is an important part of the exploration of star formation. Warm clouds
have been fairly well studied in lines of CO, CS, NH$_3$,
CH$_3$OH and many other molecules. However, the CO, CS, NH$_3$, and
CH$_3$OH  emissions are strongly affected by the contribution from
compact regions of hot gas (hot cores), where the abundances of these
and some other molecules are enhanced by several orders of magnitude
owing to grain mantle evaporation.

To eliminate the contribution from hot cores, one should observe warm
clouds in lines of  molecules that weakly emit in  such regions.
One of these molecules is methyl acetylene. Its abundance is not enhanced in
the Orion Hot Core~\citep{wang}, and, probably, it is not enhanced in
other hot cores as well. Since hot cores, which have sizes
of aproximately 0.05 pc, are much smaller than warm clouds, their
contribution to the CH$_3$CCH emission is probably negligible.

The rotational lines of methyl acetylene are grouped in
series of $J_K-(J-1)_K$ lines with closely spaced frequencies. The lines
from the same series can be observed simultaneously with the same receiver,
making it possible to obtain accurate ratios of their intensities.
Methyl acetylene observations  allow determinations of the gas kinetic
temperature. Radiative transitions between different $K$-ladders are
prohibited by the selection rule $\Delta K=0$; thus, the ratios of
different $K$-ladders' populations are determined by collisions
and depend mostly on the gas kinetic temperature. Since methyl acetylene
has a fairly small dipole moment, 0.78~D~\citep{bauer}, it is collisionally
thermalized even at density $\approx 10^4$ cm$^{-3}$. In addition, 
methyl acetylene lines appeared to be optically thin in the warm clouds
that have been already observed~\citep{wang}. These properties allow
us to assume local thermodynamic equilibrium (LTE) and utilize simple methods
of analysis, for instance, rotational diagrams, which produce fairly accurate
temperature estimates in this case. Simulations
by \citet{askne}
showed that the methyl acetylene rotational temperatures are approximately
equal to the gas kinetic temperatures.

\begin{table}
\begin{center}
\caption{Parameters of the observed methyl acetylene lines
\label{lines}
}
\vskip 5mm
\begin{tabular}{lccc}
\hline\noalign{\smallskip}
Transition    &Frequency  & S   &E$\rm_u$/h\\
              &(GHz)      &     &(K)\\
\noalign{\smallskip}
\hline\noalign{\smallskip}
$6_0-5_0$     &102547.984 &6.00 &12.315\\
$6_1-5_1$     &102546.024 &5.83 &19.300\\
$6_2-5_2$     &102540.144 &5.33 &40.258\\
$6_3-5_3$     &102530.348 &4.50 &75.181\\
$6_4-5_4$     &102516.573 &3.33 &124.060\\
$6_5-5_5$     &102499.110 &1.83 &186.885\\
$5_0-4_0$     &85457.272  &5.00 &8.209\\
$5_1-4_1$     &85455.622  &4.80 &15.196\\
$5_2-4_2$     &85450.730  &4.20 &36.153\\
$5_3-4_3$     &85442.528  &3.20 &71.076\\
$5_4-4_4$     &85431.224  &1.80 &119.956\\
\noalign{\smallskip}
\hline\noalign{\smallskip}
\end{tabular}
\end{center}
\end{table}

\section{OBSERVATIONS}

\begin{figure*}
\resizebox{0.9\linewidth}{!}{\includegraphics{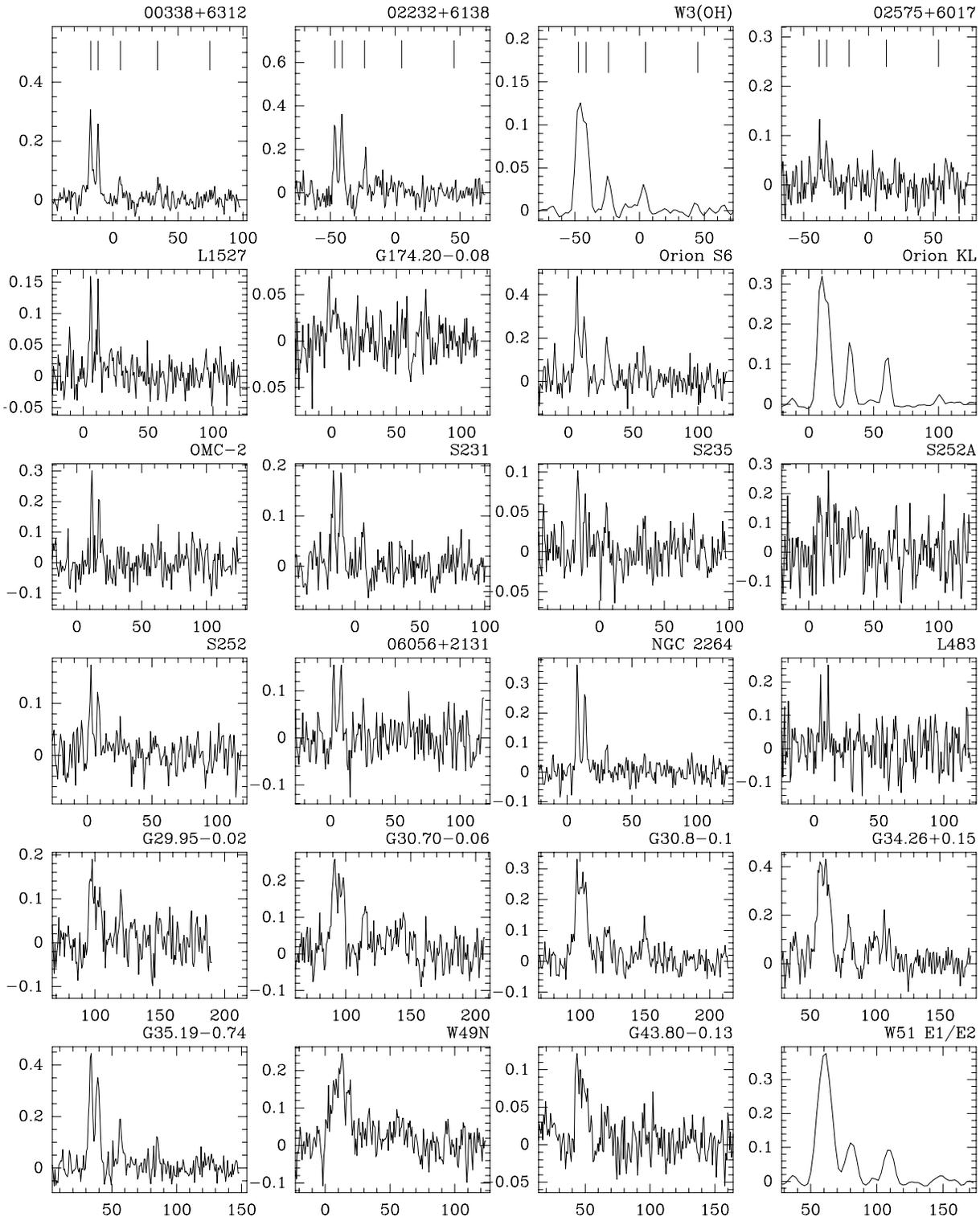}}
\caption{Spectra of the sources detected in the $6_K-5_K$ lines.
X-axis plots the LSR velocity of the $6_0-5_0$ line in km/s;
Y-axis plots the antenna temperature in Kelvins. The vertical lines
in the upper row indicate the positions of different $K$ components
($K$ values increase to the right from 0 to 4).}
\end{figure*}

\begin{figure*}
\resizebox{0.9\linewidth}{!}{\includegraphics{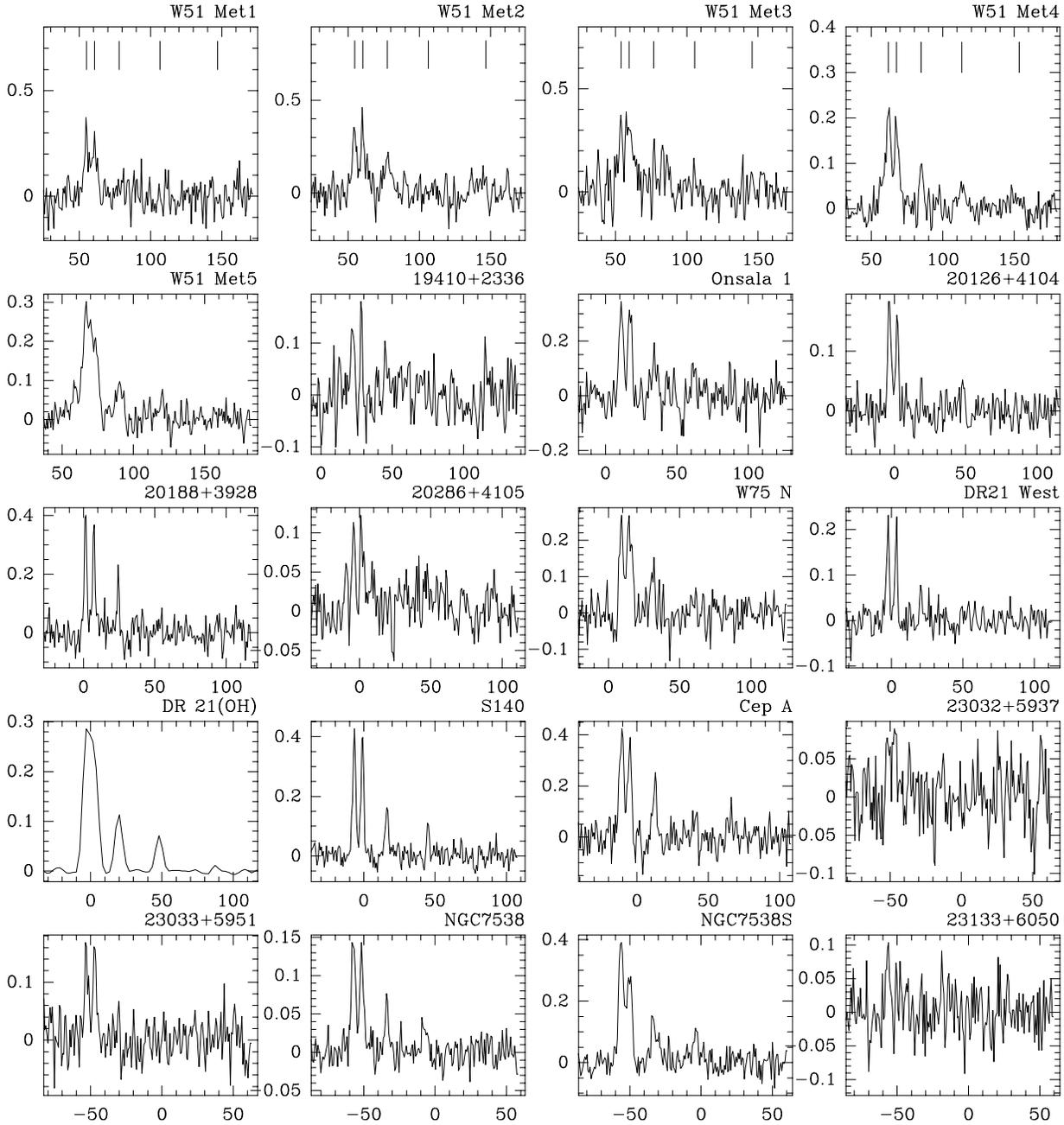}}
\caption{Spectra of the sources detected in the $6_K-5_K$ lines (continued)}
\end{figure*}

\begin{figure*}
\resizebox{0.9\linewidth}{!}{\includegraphics{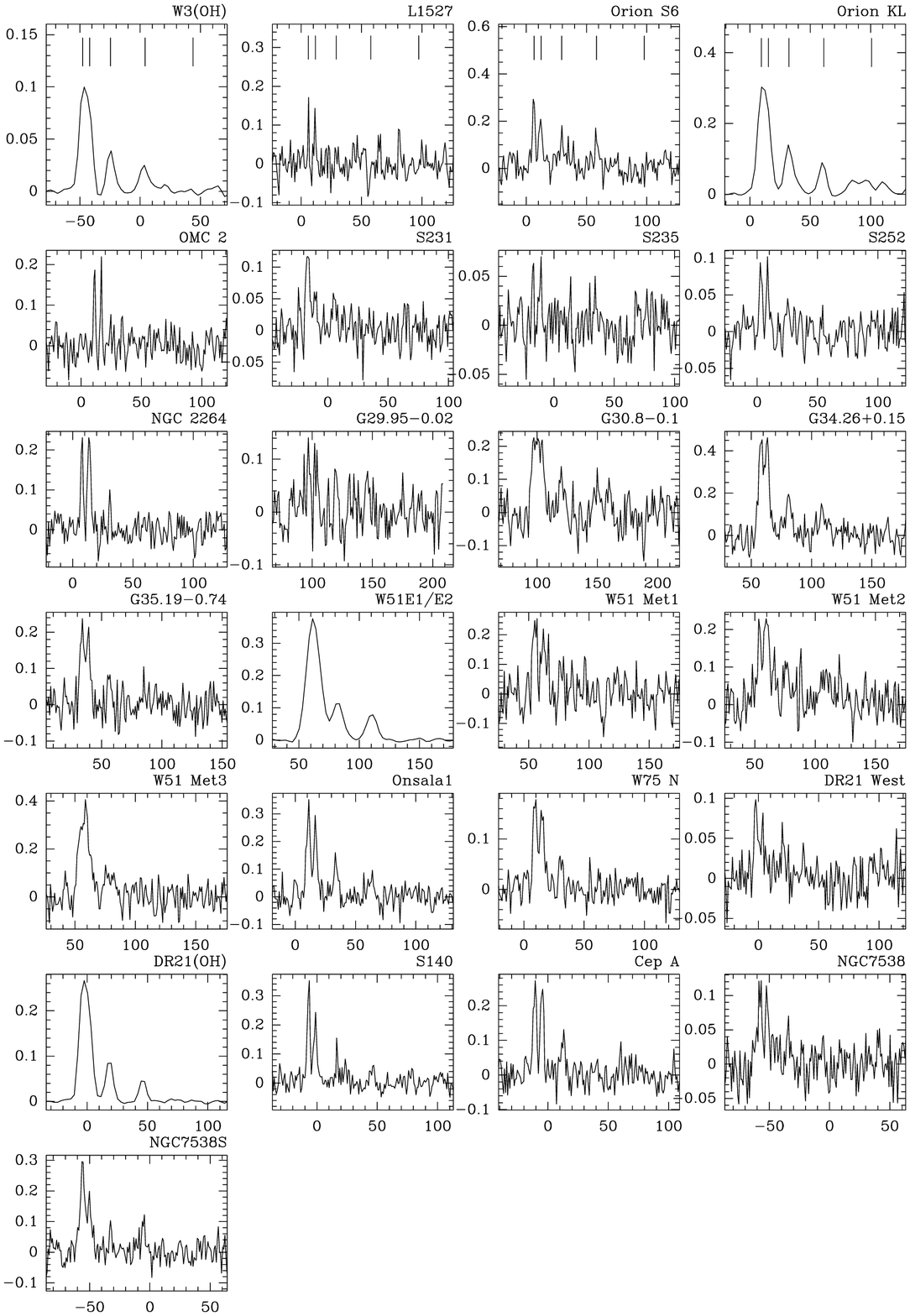}}
\caption{Spectra of the sources detected in the $5_K-4_K$ lines.
X-axis plots the LSR velocity of the $5_0-4_0$ line in km/s;
Y-axis plots the antenna temperature in Kelvins.}
\end{figure*}

The observations were carried out in May 1997\footnote{Sensitive observations
of four sources, W~3(OH), Orion~KL, W~51E1/E2, and DR~21(OH) were conducted
in 2000} with the 20-m millimeter-wave
radio telescope of the Onsala Space Observatory, Sweden. The frequency,
line strength, and upper level energy for each observed transition
are given in Table 1. Pointing accuracy was checked using observations of
SiO masers and was no worse than $5''$. The observations were performed
in the dual beam switching mode with a beam separation of $11'$ and
a switch frequency of 2 Hz. The main-beam efficiency and full-width at half
power were 0.55 and $38''$, respectively, at 102 GHz, and 0.6 and $43''$
at 85 GHz. A cooled low-noise SiS mixer was used for the observations at both
frequencies. The system noise temperature, corrected for atmospheric
absorption, rearward spillower, and radome losses, varied between 350 and
2000~K, depending on the weather conditions and source elevation. The data
were calibrated using the chopper-wheel method. The backend consisted
of two parallel filter spectrometers: a 256-channel spectrometer with 250~kHz
resolution and a 512-channel spectrometer with 1~MHz resolution. Sixty-three
sources were observed at 102 GHz. Thirty-two of these were observed at 85~GHz.
When fitting Gaussians to the lines, we assumed that different $J_K-(J-1)_K$
lines with the same $J$ have identical radial velocities and widths.
The five sources NGC 2264, G30.8-0.1, G34.26+0.15, DR 21(OH), and S140 were
mapped at 102~GHz. The mapping technique is described below.

Forty-five sources were detected at 102 GHz and twenty-five --- at 85~GHz.
The source coordinates and Gaussian parameters of the detected lines are
presented in Table 2. The spectra of the detected sources are shown in
Figs. 1, 2, and 3.

\section{DATA ANALYSIS}

The spectra in Figs. 1, 2, and 3 demonstrate that in most of
the sources only the $K=0$, $K=1$, and $K=2$ lines with low excitation
energies (Table 1) were detected, the $K=0$ and 1 lines being usually
blended. Sometimes the $K=3$ line can be distinguished from the noise.
Only in the four sources, observed in 2000 with a high sensitivity,
very weak $K=4$ lines were detected.

Figure 4 show the $6_K-5_K$ methyl cyanide spectra of three sources,
observed by \citet{kalensky2} with the same equipment and with
approximately the same sensitivity as the methyl acetylene lines.
Methyl cyanide (CH$_3$CN), like methyl acetylene, is a symmetric top molecule
and its spectrum also consists of $J_K-(J-1)_K$ groups. The methyl
acetylene and methyl cyanide spectra of NGC~7538S look alike (there are no
lines with $K>3$ in both spectra) and the radial velocities are the same;
therefore one can assume that the emission of both molecules arise in
the same region. In W~3(OH) and G34.26+0.15, the contribution from hot cores
to the methyl cyanide emission is significant~\citep{kalensky2}. In these objects,
high energy lines are clearly detectable in the methyl cyanide spectra,
but they are not noticeable in the methyl acetylene spectra. The high-energy
CH$_3$CN lines appear in the spectra of these sources due to the contribution
from hot cores~\citep{kalensky2}, and the absence of detectable emission in analogous methyl
acetylene lines suggests that the contribution from hot cores to the CH$_3$CCH
emission is insignificant. High-energy CH$_3$CCH lines are either weak
or not detected in all the observed sources. Since many of them
harbor hot cores, it is reasonably to assume that the contribution from
hot cores to the methyl acetylene emission is typically small.

We built rotational diagrams for 40 sources at 102~GHz and for 23 sources
at 85~GHz (Fig. 6). The temperatures, derived from the rotational diagrams,
fall in the range from 20 to 60 K, typical for warm clouds,
and within errors agree with those derived from the observations of ammonia,
methanol, and methyl cyanide~\citep{kalensky1, kalensky2, mauersberger,
churchwell, molinari, wouterloot, wilson, schreyer}.
Methyl acetylene column densities vary between $4\times 10^{13}$~cm$^{-2}$ and $1.2\times 10^{15}$~cm$^{-2}$,
in most of the sources being included in a much narrower range,
$1-5\times 10^{14}$~cm$^{-2}$.

The observations of W~3(OH), DR~21(OH), Orion~KL, and W~51E1/E2, which
resulted in a detection of a sufficient number of lines (up to $K=4$), were
analysed not assuming that the lines are optically thin. We calculated
the ratios of line brightness temperatures for different sets of gas 
temperatures and CH$_3$CCH column densities assuming that the CH$_3$CCH level
populations are thermalized. The temperature varied
between 10 and 200~K, column density~---~between $10^{11}$ and
$10^{15}$~cm$^{-2}$, the sources were assumed to be uniform. Then we selected
the models that agreed with the obsevations according to the $\chi^2$
criterion. The results are presented in Fig. 5, which shows that both
optically thin models and models with the opacities $\tau$ of the $K=0$
and 1 lines about unity may match the observations. In the latter case,
the intensities of the lines proved to be approximately an order of
magnitude higher than the observed intensities; hence, the models with
$\tau\approx 1$ are possible only if the CH$_3$CCH emission is significantly
diluted. Note that the gas temperatures, derived from the models with
$\tau \approx 1$ is only slightly lower than those derived from the rotational
diagrams. Below we consider that the methyl acetylene lines are optically
thin.

In the last column of Table 3 we present the methyl acetylene abundances
$N_{\rm CH_3CCH}/N_{\rm H_2}$, which were determined employing $^{13}$CO data
\citep{kalensky2}. The observatons of the 1~--~0 $^{13}$CO line at 110~GHz have also been performed
at Onsala. The $^{13}$CO and H$_2$ column densities were calculated
assuming that the 1~--~0 $^{13}$CO lines are optically thin, the gas
temperatures are equal to the CH$_3$CCH rotational temperatures, and
the $^{13}$CO abundance is equal to $1.7\times 10^{-6}$~\citep{lucas&liszt}.

\subsection{THE TECHNIQUE OF MAPPING}

\begin{table*}
\caption{Gaussian parameters for detected CH$_3$CCH lines with $1\sigma$
errors. For each source the parameters of the 102~GHz lines are presented
in the upper row, and those of the 85~GHz lines in the lower row.}
\footnotesize
\medskip
\begin{tabular}{|lllllcccc|}
\hline\noalign{\smallskip}
Source       &R.A.$_{\rm 1950}$ & \multicolumn{5}{c}{$\int T_A^*dV$ (K$\cdot$km s$^{-1}$)}     & V$_{LSR}$   & FWHM     \\
&DEC.$_{\rm 1950}$ &K=0&K=1&K=2&K=3&K=4&(km s$^{-1}$)&(km s$^{-1}$)\\
\noalign{\smallskip}
\hline\noalign{\smallskip}
00338+6312& 00 33 53.3&0.81(0.05)&0.66(0.04)&0.22(0.04)&0.15(0.04)&$<$0.12  &-17.33(0.06)&2.74(0.1)\\
          & 63 12 33.0& not observed     &          &          &          &     &            &         \\
02232+6138&02 23 13.4&0.93(0.07)&1.06(0.08)&0.35(0.08)&$<$0.21   &$<$0.21  &-46.56(0.08)&2.79(0.1)\\
          & 61 38 44.8& not observed      &          &          &          &     &            &         \\
W 3(OH)   &02 23 17.3&0.74(0.02)&0.55(0.02)&0.24(0.02)&0.18(0.02)    &0.04(0.02) &-47.097(0.09)&5.43(0.10)\\
          &61 38 58.0&0.64(0.02)&0.48(0.02)&0.28(0.01)&0.18(0.01)    &           &-47.679(0.11)&6.74(0.12)\\
02575+6017&02 57 35.6&0.21(0.04)&0.17(0.04)&$<$0.12)    &$<$0.12     &$<$0.15&-37.96(0.15)&1.75(0.24)\\
          &60 17 22.6&   not observed    &          &          &            &         &            &          \\
L 1527    &04 36 49.3&0.22(0.03)&0.18(0.03)&$<$0.05     &          &     & 5.78(0.07) &1.21(0.10)\\
          &25 57 16.0&0.16(0.05)&0.19(0.07)&$<$0.09     &          &     & 5.97(0.08) &0.74(0.14)\\
G174.20   &05 27 32.2&0.16(0.04)&0.13(0.05)&$<$0.12     &         &     & -2.39(0.35)&3.01(0.58)\\
          &33 45 52.0&  not observed    &           &           &         &     &             &         \\
Orion S6  &05 32 44.8&1.42(0.10)&0.95(0.09)&0.60(0.09)&$<$0.25     &     & 6.68(0.10) &3.22(0.13)\\
         &-5 26 00.0&0.66(0.09)&0.40(0.08)&0.44(0.08)&$<$0.24     &     & 6.48(0.11) &2.26(0.16)\\
Orion KL &05 32 47.0&1.93(0.04)&1.46(0.04)&0.99(0.04)&0.79(0.04)  &$<$0.13&9.13(0.06)&5.61(0.07)\\
         &-5 24 20.0&2.06(0.10)&1.41(0.10)&1.08(0.06)&0.70(0.06)  &     & 9.43(0.16) &7.53(0.18)\\
OMC 2    &05 32 59.9&0.53(0.07)&0.39(0.06)&$<$0.18     &          &     & 11.46(0.08)&1.47(0.11)\\
         &-5 11 29.0&0.34(0.05)&0.30(0.05)&$<$0.12     &          &     & 11.17(0.07)&1.31(0.11)\\
S 231    &05 35 51.3&0.45(0.05)&0.45(0.05)&0.20(0.05)&$<$0.15     &     &-16.46(0.10)&2.43(0.15)\\
         &35 44 16.0&0.45(0.05)&0.18(0.05)&0.16(0.05)&$<$0.15     &     &-16.56(0.21)&3.38(0.22)\\
S 235    &05 37 31.8&0.21(0.03)&0.11(0.03)&0.13(0.03)&$<$0.09     &     &-16.90(0.13)&2.08(0.16)\\
         &35 40 18.0 &0.14(0.03)&0.14(0.03)&$<$0.9    &            &     &            &          \\
S 252A   &06 05 36.5&0.52(0.14)&0.47(0.14)&$<$0.18     &          &     & 8.77(0.33) &2.78(0.30)\\
         &20 39 34.0& not observed&          &          &          &     &            &          \\
S 252    &06 05 53.7&0.31(0.05)&0.25(0.05)&0.10(0.04)&$<$0.12     &$<$0.12& 2.74(0.13) &2.11(0.23)\\
         &21 39 09.0&0.29(0.04)&0.27(0.04)&$<$0.12     &          &     & 2.94(0.15) &2.78(0.21)\\
06056+2131&06 05 41.0&0.40(0.06)&0.45(0.07)&0.18(0.06)&$<$0.15     &$<$0.15& 2.61(0.12) &2.21(0.21)\\
          &21 31 32.2& not observed      &          &          &          &         &            &          \\
NGC 2264  &06 38 24.9&0.75(0.05)&0.63(0.05)&0.18(0.04)&$<$0.12     &     & 8.06(0.05) &2.10(0.08)\\
          &09 32 28.0&0.57(0.05)&0.68(0.05)&0.19(0.05)&$<$0.15     &     & 7.97(0.08) &2.47(0.11)\\
L 483     &18 14 50.6&0.22(0.07)&0.25(0.06)&$<$0.10     &$<$0.10     &$<$0.10& 5.11(0.14) &0.96(0.12)\\
          &-04 40 49.0&not observed&       &          &            &       &            &          \\
G29.95-0.02&18 43 27.1&0.71(0.08)&0.47(0.08)&0.31(0.08)&$<$0.21     &$<$0.15&97.35(0.21) &3.78(0.21)\\
           &-02 42 36.0&0.44(0.08)&0.44(0.08)&0.30(0.07)&$<$0.21     &       &            &          \\
G30.70-0.06&18 44 58.9&1.15(0.09)&0.95(0.09)&0.44(0.08)&0.33(0.08)&$<$0.24&91.27(0.19) &4.53(0.19)\\
           &-02 04 27.0&not observed       &          &          &          &       &            &          \\
G30.8-0.1  &18 45 11.0&1.43(0.08)&1.36(0.08)&0.58(0.08)&0.52(0.07)&$<$0.6 &98.00(0.00) &5.35(0.20)\\
           &-01 57 57.0&1.49(0.15)&1.42(0.15)&0.85(0.14)&0.51(0.13)&$<$0.09&97.83(0.31) &6.01(0.37)\\
G34.26+0.15&18 50 46.1&2.32(0.12)&2.42(0.12)&1.27(0.10)&0.49(0.10)&$<$0.3 &57.64(0.15) &6.02(0.17)\\
           &01 11 12.0&2.26(0.12)&2.16(0.12)&1.03(0.11)&0.54(0.11)&$<$0.3 &57.91(0.13) &4.75(0.15)\\
G35.19-0.74&18 55 40.8&1.52(0.07)&1.30(0.07)&0.62(0.03)&0.34(0.04)&$<$0.18&33.61(0.01) &3.37(0.08)\\
           &01 36 30.0&0.98(0.09)&0.88(0.09)&0.43(0.08)&$<$0.24     &     &33.63(0.01) &4.16(0.19)\\
W 49N      &19 07 49.9&0.79(0.88)&2.41(0.84)&0.73(0.14)&0.94(0.15)&$<$0.42& 7.14(1.69) &13.49(0.77)\\
           &09 01 14.0& not observed      &          &          &          &       &            &           \\
G43.80-0.13&19 09 30.8&0.57(0.06)&0.44(0.05)&0.28(0.05)&$<$0.12     &$<$0.12&44.41(0.28) &5.04(0.28)\\
           &09 30 47.0& not observed      &          &          &           &        &            &          \\
W 51E1/E2  &19 21  26.2&2.70(0.19)&2.56(0.18)&1.49(0.07)&1.18(0.08)&0.23(0.07)&57.79(0.22) &9.93(0.16)\\
           &14 24 43.0&2.89(0.36) &2.16(0.34)&1.26(0.10)&0.97(0.10)&          &59.85(0.43) &10.73(0.26)\\
W 51 MET1  &19 21 26.2&1.17(0.13)&0.97(0.13)&$<$0.36     &$<$0.27     &$<$.27 &55.03(0.22) &3.92(0.24)\\
           &14 23 32.0&1.31(0.15)&0.93(0.14)&$<$0.39     &            &       &            &          \\
W 51 MET2  &19 21 28.8&1.52(0.14)&1.54(0.15)&0.98(0.15)&$<$0.39     &$<$0.39&54.64(0.19) &4.36(0.27)\\
           &14 23 47.0&1.24(0.35)&1.15(0.30)&0.62(0.16)&0.64(0.10)  &$<$0.21&55.53(0.87) &9.10(0.80)\\
W 51 MET3 &19 21 27.5&1.30(0.15)&1.50(0.15)&0.45(0.15)&$<$0.45     &     &53.84(0.21) &3.50(0.00)\\
          &14 23 52.0&2.42(1.41)&1.41(1.49)&0.85(0.20)&$<$0.48     &     &55.91(2.89) &7.88(1.15)\\
\noalign{\smallskip}
\hline\noalign{\smallskip}
\end{tabular}
\end{table*}

\addtocounter{table}{-1}
\begin{table*}
\caption{(Continued)}
\footnotesize
\medskip
\begin{tabular}{|lllllcccc|}
\hline\noalign{\smallskip}
Source    &R.A.$_{\rm 1950}$ &\multicolumn{5}{c}{$\int T_A^*dV$ (K$\cdot$km s$^{-1}$)}     & V$_{LSR}$            & FWHM     \\
          &DEC.$_{\rm 1950}$ &K=0&K=1&K=2&K=3&K=4                                     &(km$\cdot$ s$^{-1}$)&(km $\cdot$ s$^{-1}$)\\
\noalign{\smallskip}
\hline\noalign{\smallskip}
W 51 MET4 &19 21 25.6&1.06(0.05)&0.81(0.05)&0.25(0.05)&0.30(0.05)&$<$0.15&61.77(0.12)&4.33(0.15)\\
          &14 25 41.0& not observed      &          &           &         &           &           &        \\
W 51 MET5 &19 21 20.5&2.25(0.05)&1.13(0.05)&0.75(0.05)&0.39(0.05)&$<$0.15&67.08(0.73) &9.34(0.73)\\
          &14 24 12.0& not observed     &          &          &          &       &            &           \\
19410+2336&19 41 04.2&0.34(0.07)&0.34(0.05)&0.18(0.06)&$<$0.18     &$<$0.18&22.55(0.13) &1.98(0.18)\\
          &23 36 42.0& not observed     &          &          &            &       &            &         \\
Onsala 1  &20 08 09.9&1.24(0.11)&1.18(0.10)&0.58(0.10)&0.39(0.10)&$<$0.27&11.17(0.12) &3.42(0.13)\\
          &31 22 42.0&1.24(0.07)&0.98(0.07)&0.53(0.07)&0.31(0.07)&$<$0.21&11.05(0.09) &3.60(0.11)\\
20126+4104&20 12 40.96&0.55(0.03)&0.41(0.02)&0.15(0.02)&0.11(0.02)&$<$0.06&-3.64(0.04) &2.46(0.07)\\
          &41 04 20.6& not observed     &          &          &          &       &            &         \\
20188+3928&20 18 50.6&1.02(0.09)&0.87(0.08)&0.43(0.08)&0.35(0.08)&$<$0.24  &1.46(0.07)  &2.11(0.12)\\
          &39 28 18.6& not observed      &         &           &           &      &            &          \\
20286+4105&20 28 40.6&0.45(0.05)&0.45(0.05)&$<$0.15     &$<$0.15     &$<$0.15&-4.04(0.17) &3.36(0.20)\\
          &41 05 38.1& not observed     &          &           &           &         &            &          \\
W 75N     &20 36 50.4&0.96(0.08)&1.05(0.08)&0.44(0.08)&$<$0.24     &     &9.03(0.13)  &3.71(0.14)\\
          &42 27 23.0&0.68(0.05)&0.60(0.05)&0.16(0.05)&$<$0.12     &     &9.34(0.13)  &3.67(0.13)\\
DR 21 West&20 37 07.6&0.50(0.04)&0.45(0.04)&0.16(0.03)&$<$0.12     &     &-2.39(0.06) &2.04(0.09)\\
           &42 08 46.0&0.32(0.05)&0.21(0.05)&0.14(0.04)&$<$0.12     &     &-2.16(0.23) &3.21(0.33)\\
DR 21(OH)  &20 37 13.8&1.72(0.02)&1.45(0.02)&0.72(0.01)&0.43(0.01)&0.06(0.01)&-3.00(0.03) &5.87(0.03)\\
           &42 12 13.0&1.59(0.03)&1.30(0.03)&0.60(0.02)&0.32(0.02)&$<$0.22&-4.11(0.08) &6.46(0.08)\\
S 140      &22 17 41.2&1.09(0.05)&1.07(0.05)&0.44(0.05)&0.16(0.05)&$<$0.15&-6.62(0.04) &2.46(0.06)\\
           &63 03 43.0&1.00(0.06)&0.69(0.06)&0.29(0.05)&0.17(0.05)&$<$0.15&-6.74(0.07) &2.88(0.11)\\
Cep A      &22 54 19.2&1.56(0.10)&1.26(0.10)&0.80(0.10)&0.34(0.09)&$<$0.27&-10.70(0.09)&3.35(0.11)\\
           &61 45 47.0&0.89(0.07)&0.88(0.07)&0.43(0.07)&$<$0.21     &     &-10.58(0.10)&3.19(0.11)\\
23032+5937&23 03 16.9&0.23(0.07)&0.29(0.07)&$<$0.18     &$<$0.18     &$<$0.18&-51.91(0.34)&2.99(0.43)\\
          &59 37 38.9& not observed     &          &            &             &      &            &         \\
23033+5951&23 03 19.7&0.48(0.06)&0.49(0.06)&$<$0.20     &$<$0.20     &$<$0.20&-52.94(0.13)&2.83(0.19)\\
          &59 51 55.0& not observed     &          &            &            &       &            &           \\
NGC 7538  &23 11 36.6&0.55(0.03)&0.46(0.03)&0.21(0.03)&0.11(0.03)&$<$0.09&-57.19(0.08)&3.36(0.10)\\
          &61 11 50.0&0.34(0.05)&0.24(0.05)&0.12(0.04)&$<$0.15     &     &-57.53(0.18)&2.89(0.22)\\
NGC 7538S &23 11 36.1&1.73(0.07)&1.27(0.06)&0.60(0.06)&0.40(0.06)&$<$0.18&-55.99(0.10)&4.09(0.09)\\
          &61 10 30.0&1.15(0.07)&0.67(0.07)&0.25(0.07)&0..32(0.07)&$<$0.20&-55.76(0.13)&3.50(0.00)\\
23133+6050&23 13 21.5&0.24(0.07)&0.12(0.05)&$<$0.15&$<$0.15   &$<$0.15&-56.30(0.19)&1.90(0.39)\\
           &60 50 45.6&             &          &       &          &       &            &          \\
\hline\noalign{\smallskip}
\end{tabular}
\end{table*}

\begin{figure*}
\begin{center}
\resizebox{0.5\linewidth}{!}{\includegraphics{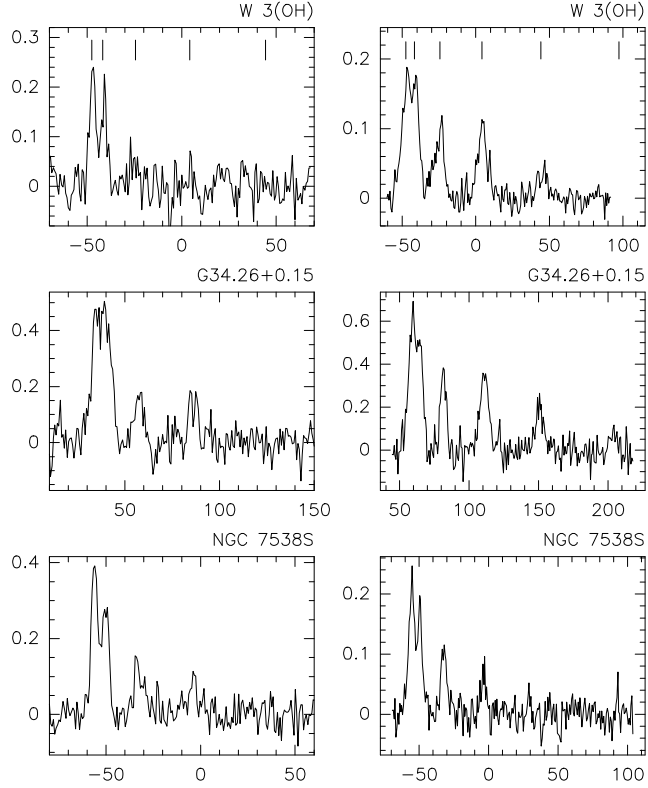}}
\caption{Comparison of CH$_3$CCH and CH$_3$CN spectra (left and right
columns, respectively).}
\end{center}
\end{figure*}

\begin{figure*}
\begin{center}
\resizebox{\linewidth}{!}{\includegraphics{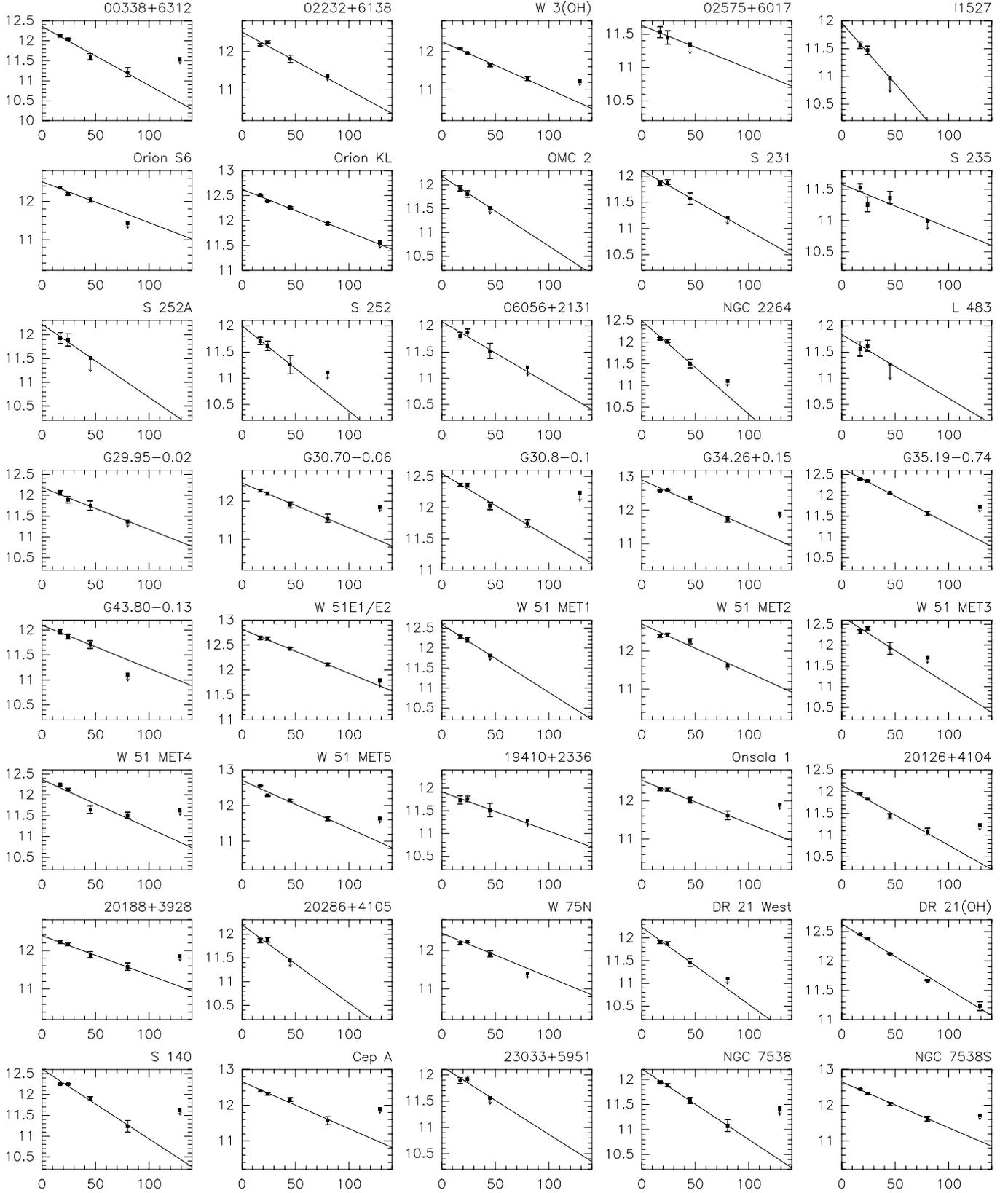}}
\caption{Rotational diagrams for the $6_K-5_K$ lines. X-axis plots
the upper level energies divided by the Boltzman constant $E_u/k$,
Y-axis plots the upper level populations divided by statistical veights
$\frac{\lg(3kW)}{8\pi^3\nu_0S\mu^{2}_{ul}g_Ig_k}$ values. The arrows
denote the upper level populations of undetected lines at 3~$\sigma$ level.}
\end{center}
\end{figure*}

\begin{figure*}
\begin{minipage}[b]{.49\linewidth}
\includegraphics[width=\linewidth]{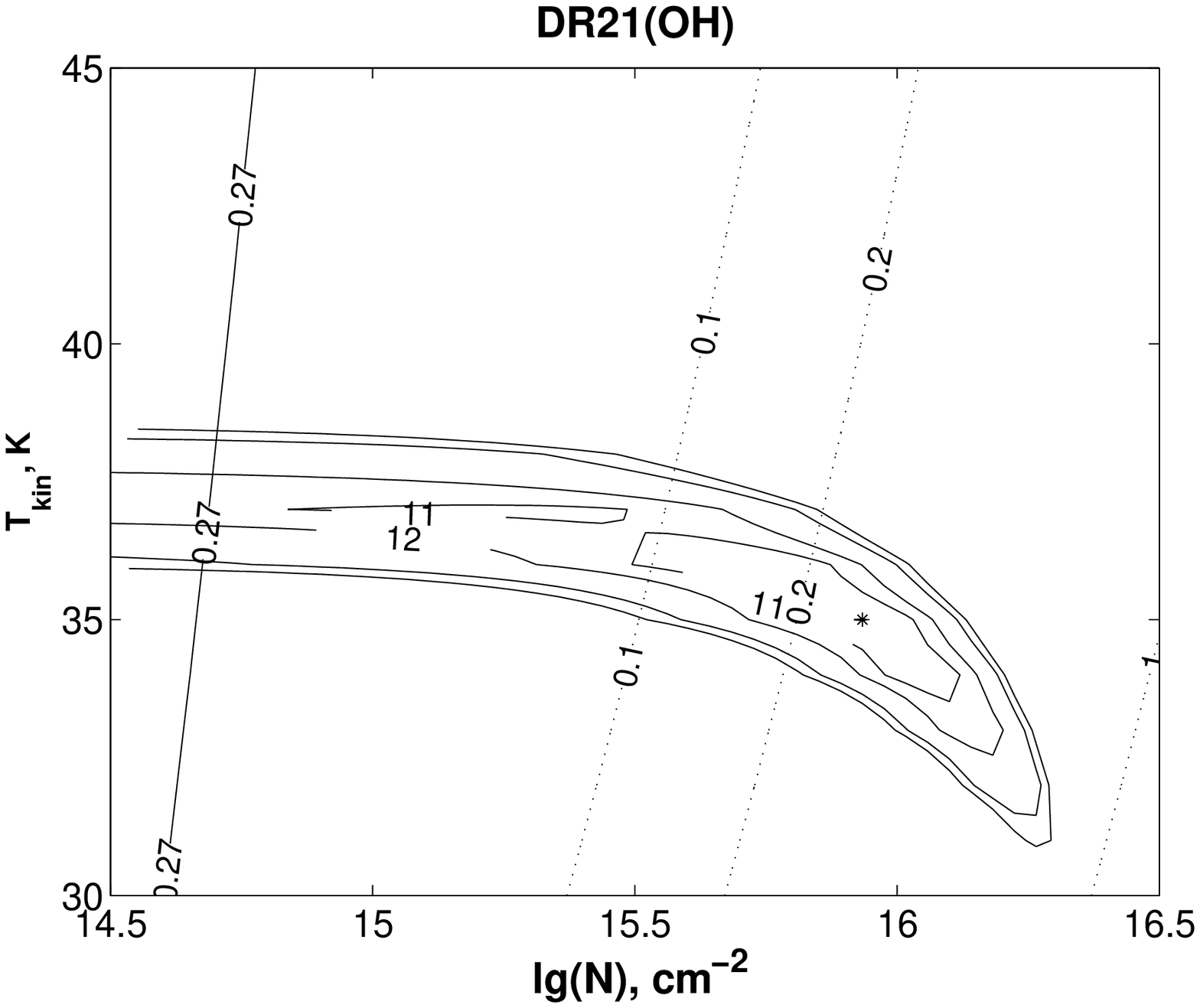}
\end{minipage}
\hfill
\begin{minipage}[b]{.49\linewidth}
\includegraphics[width=\linewidth]{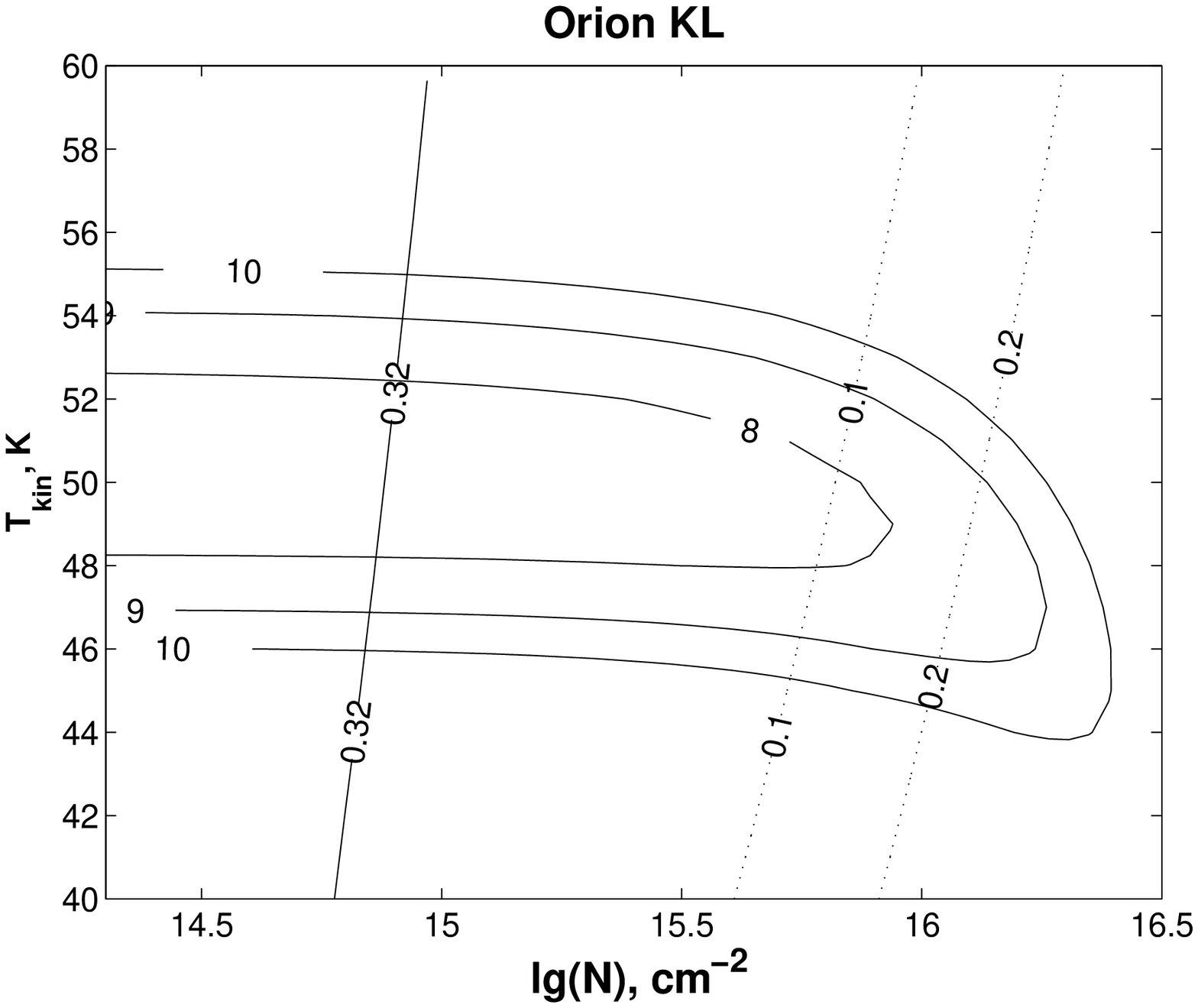}
\end{minipage}
\vskip 5mm
\begin{minipage}[b]{.49\linewidth}
\includegraphics[width=\linewidth]{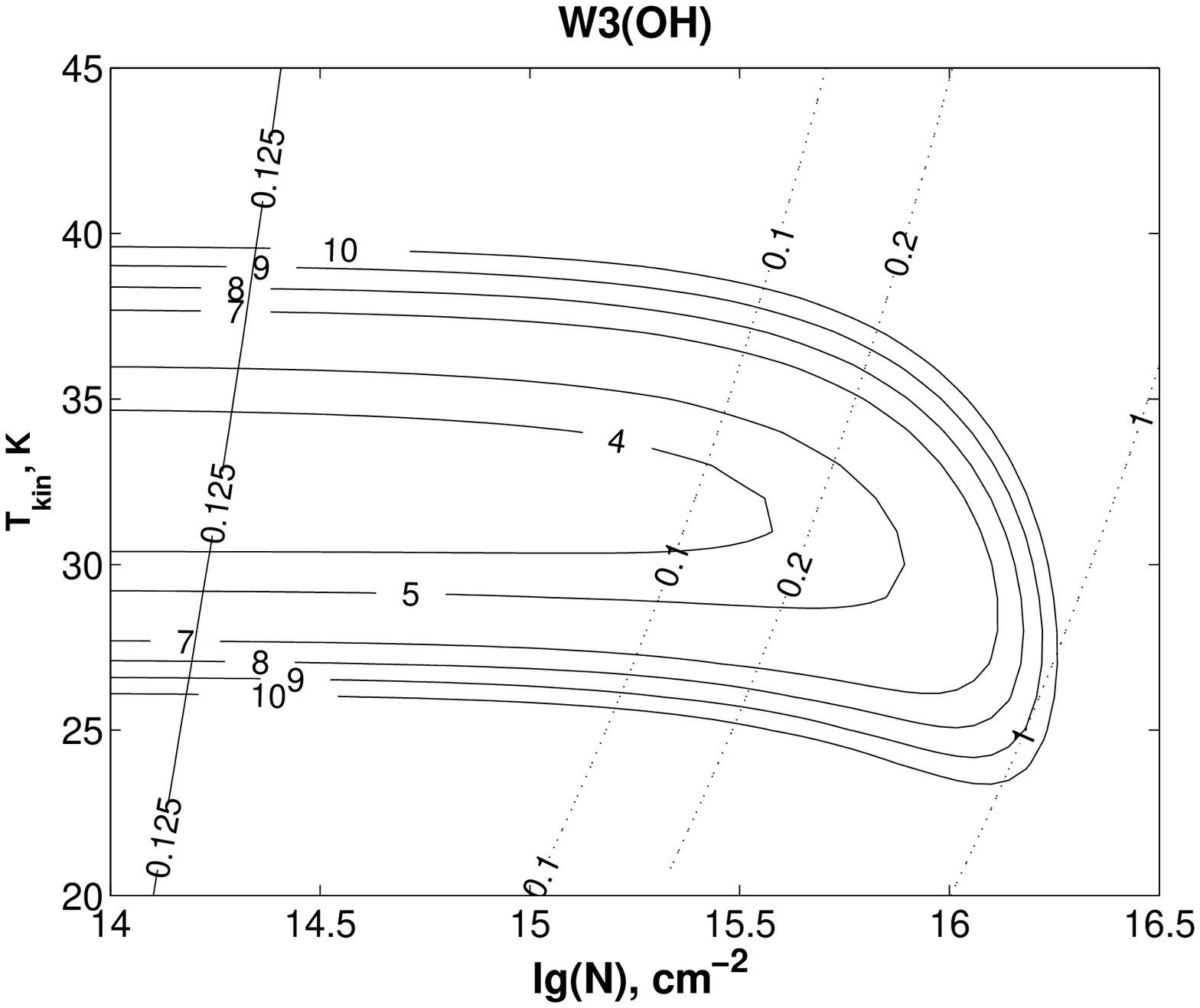}
\end{minipage}
\hfill
\begin{minipage}[b]{.49\linewidth}
\flushright\includegraphics[width=\linewidth]{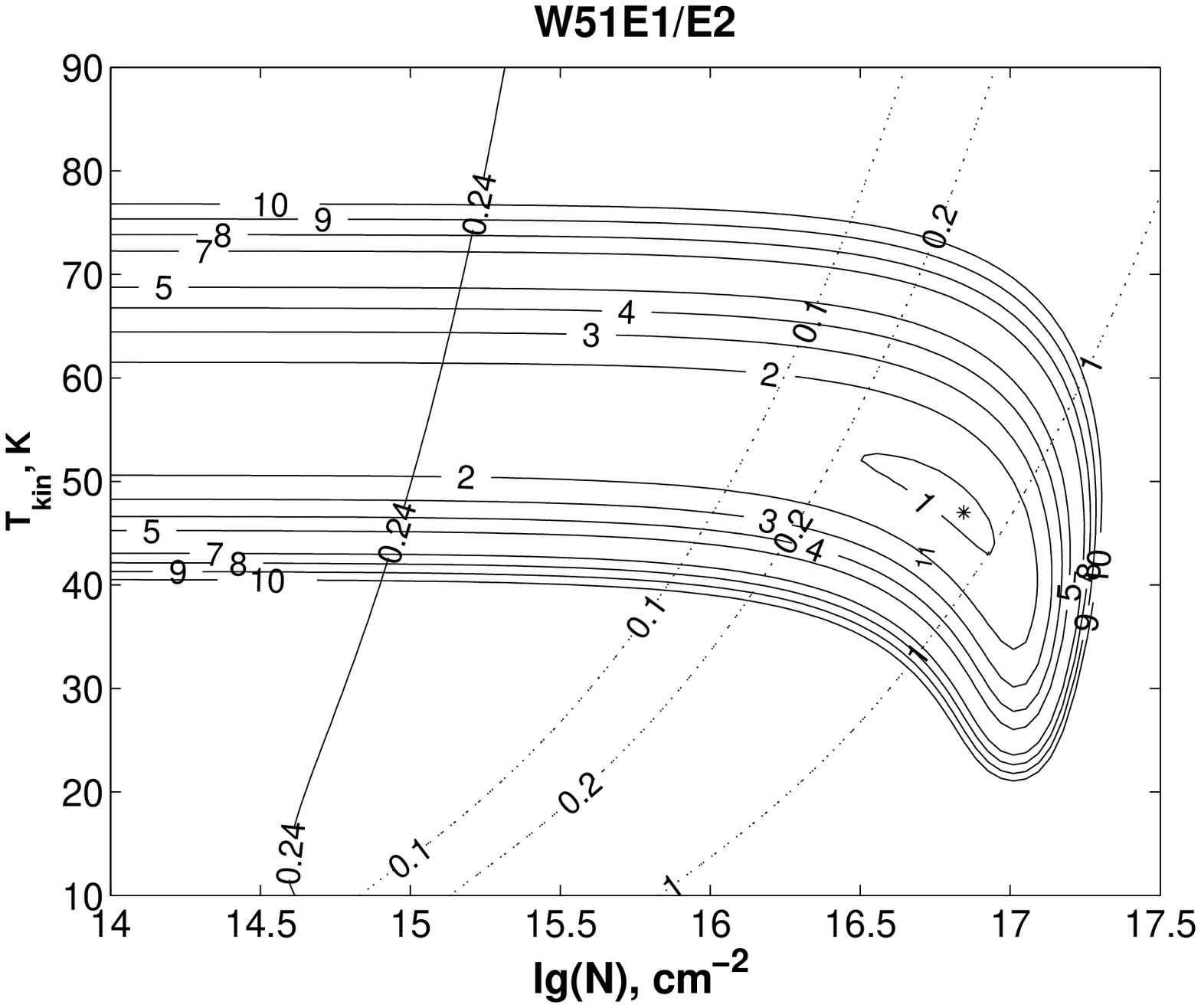}
\end{minipage}
\caption{Parameters of the sources determined by the $\chi^2$ method.
The contours show the derived $\chi^2$ values, the oblique solid
lines correspond to the parameter sets that produce the brightness
temperature of the $6_0-5_0$ line equal to the observed one.
The dotted lines correspond to the parameter sets that provide
the optical depths of the $6_0-5_0$ line equal to 0.1, 0.2, and 1.}
\end{figure*}

\begin{table*}
\begin{center}
\small
\caption{Rotational temperatures and column densities determined from
the methyl acetylene observations. The CH$_3$CCH abundances were
derived from the column densities that were determined using the $6_K-5_K$
lines.}
\vskip 5pt
\hskip -5mm
\small
\begin{tabular}{|l|c|c|c|c|c|}
\hline\noalign{\smallskip}
            &\multicolumn{2}{c|}{102 GHz}&\multicolumn{2}{c|}{85 GHz}&\\
\cline{2-6}
Source      &$T_{rot}$&$N_{\rm CH_{3}CCH}$&$T_{rot}$&$N_{\rm CH_{3}CCH}$&CH$_3$CCH abundance\\
            &  (K)    &$\times10^{14}$    &   (K)      &$\times10^{14}$    &$\times10^{-9}$  \\
\noalign{\smallskip}
\hline\noalign{\smallskip}
00338+6312      & 29   (1)     &1.9 (0.2)    &            &                                       &\\
02232+6138      & 29   (3)     &2.6 (0.5)    &            &                                       &\\
W 3(OH)         & 35   (0.7)   &2.0 (0.1)    &42.5 (2)    &2.6 (0.2)                               &2.3  \\
02575+6017      & $<$67 (41)   &$<$1.2 (1.2) &            &                                       &\\
L 1527          & $<$20 (3)    &$<$0.4 (0.1) &            &                                       &\\
ORI S6          & 41   (6)     &4.4 (1.0)    & $<$65  (22)&$<4.5$ (2.3)                            &5.2 \\
Orion KL        & 51   (1)     &8.0 (0.3)    & 52  (4)    &10.5 (1.3)                             & \\
OMC 2           &$<$30 (6)     &1.3 (0.4)    & $<$34 (15) &$<1.1$ (0.8)                            &4.1\\
S 231           & 38   (8)     &1.5 (0.5)    & 29    (11) &1.1 (0.7)                              &4.8 \\
S 235           & $<$61  (11)  &$<1.0$ (0.3) &            &                                       &1.1 \\
S 252A          & $<$28   (10) &$<1.3$ (0.9) &            &                                       &\\
S 252           & 27   (6)     &0.7 (0.3)    &$<$42  (22) &$<1.3$ (1.0)                            &2.1 \\
06056+2131      & 36   (10)    &1.4 (0.6)    &            &                                       &\\
NGC 2264        & 20   (1)     &1.5 (0.2)    & 46    (20) &3.1 (2.0)                              &3.9 \\
L483            &$<$36 (20)    &0.8 (0.7)    &            &                                       &                   \\
G29.95-0.02     & 43   (10)    &2.3 (0.9)    & 67    (25) &3.5 (2.1)                              &2.0 \\
G30.70-0.06     & 37   (2)     &3.5 (0.4)    &            &                                       &                   \\
G30.8-0.1       & 42   (2)     &5.1 (0.5)    & 50    (10) &8.2 (2.6)                              &5.7\\
G34.26+0.15     & 31   (1)     &7.2 (0.5)    & 40    (4)  &9.3 (1.7)                              &16 \\
G35.19-0.74     & 32   (1)     &4.3 (0.3)    & 44    (14) &4.4 (2.3)                              &8.5 \\
G43.80-0.13     & 50   (13)    &2.3 (1.0)    &            &                                       &\\
W 51E1/E2       & 49  (3.5)    &12 (1.5)     & 48    (5.5)&13.3 (2.7)                             &3.6\\
W 51 MET1       & $<$26  (4)   &$<2.8$ (1.6) & $<$26 (9)  &$<3.1$ (1.7)                           &\\
W 51 MET2       & $<$34   (2)  &$<6.5$ (2.2) & 69    (19) &9.4 (4.6)                              &\\
W 51 MET3       & 26   (4)     &3.5 (1.0)    & 34     (25)&7.4 (9.8)                              &2.5\\
W 51 MET4       & 37   (1)     &2.8 (0.2)   &            &                                       &\\
W 51 MET5       & 32   (0.5)   &4.9 (0.2)   &            &                                       &\\
19410+2336      & 40   (25)    &1.5 (1.3)   &            &                                       &\\
Onsala 1        & 39   (3)     &4.2 (0.6)   &39      (5) &4.7 (0.9) &6.1 \\
20126+4104      & 31   (1)     &1.3 (0.1)   &            &                                       &\\
20188+3928      & $<$42 (3)    &$<3.6$ (0.5)&            &                                       &\\
20286+4105      & $<$26 (3.7)  &$<1.1$ (0.3)&            &                                       &\\
W 75N           & 38   (6)     &3.4 (0.9)   &  29   (8)  &1.9 (0.9) &24\\
DR 21 West      & 26   (3)     &1.2 (0.2)   & 41    (19) &1.2 (0.9) &2.6 \\
DR 21 (OH)      & 39   (0.2)   &5.4 (0.1)   & 34    (1)  &5.2 (0.3) &4.5 \\
S 140           & 26    (1)    &2.8 (0.2)   & 29    (3)  &2.6 (0.5) &4.5 \\
Cep A           & 33   (1)     &4.6 (0.4)   & 53    (18) &5.2 (2.8) &5.7 \\
23033+5951      & $<$34  (7)   &$<1.5$ (0.5)&            &                                       &\\
NGC 7538        & 31   (1)     &1.5 (0.1)   &  32    (13)&1.0 (0.7) &1.9 \\
NGC 7538S       & 34   (1)     &4.6 (0.3)   &  37    (5) &3.8 (0.8) &5.2 \\
\noalign{\smallskip}
\hline\noalign{\smallskip}
\end{tabular}
\end{center}
\end{table*}

We mapped five methyl acetylene sources at 102~GHz: NGC~2264, G30.8-0.1,
G34.26+0.15, DR~21(OH), and S~140. In each of them we observed 50~--~100
positions with 10 and 20~arcsec spacings in right ascention and
declination. For each source we tried to expand the observed area until
at the edges the emission vanishes, but due to the lack of time we could
not observe sufficiently large areas towards G30.8~--~0.1 and DR~21(OH).
Therefore both the maps of these sources and the source parameters, derived
using the maps, are unreliable.

Since the beam was fairly large (about $40''$), and the signal-to-noise
ratio was often fairly low, the obtained images were heavily smoothed
by the beam and distorted by the noise. Therefore we reconstructed the images
with the method of maximum entropy.

The reconstruction of an image, distorted by the noise and
smoothed by the beam is an incorrect problem, since high spatial
frequencies are lost due to the limited spatial resolution, and there are
no ways to restore them (in other words, an infinite number of
''initial'' images match the observed smooth image). The method of maximum
entropy select from the multitude of acceptable images the one that
has the maximal Shennon entropy

\begin{equation}
E(f)=-\sum_{i=1}^N f_i log f_i
\end{equation}
where $f_i$ is the intensity of an element of the image. We used the variant
of the method, described by~\citet{wilchek}, except
the computational algorithm to solve the optimization problem.
The evolutionary algorithm, employed here, is described
by~\citet{promislov}. The latter algorithm is more expensive computationally
than that from~\citet{wilchek}, but in the case of a poor signal-to-noise ratio or
insufficient grid spacing this algorithm may converge even if the algorithm
from~\citet{wilchek} fails. Nevertheless, among the observed lines only the blends of
the $6_0-5_0$ and $5_0-4_0$ lines have sufficiently high signal-to-noise
ratios to make the mapping possible. We made the maps of the intensity
integrated over all the channels, occupied by the blends.

\begin{figure*}
\vskip 10mm
\begin{center}
\resizebox{0.7\linewidth}{!}{\includegraphics{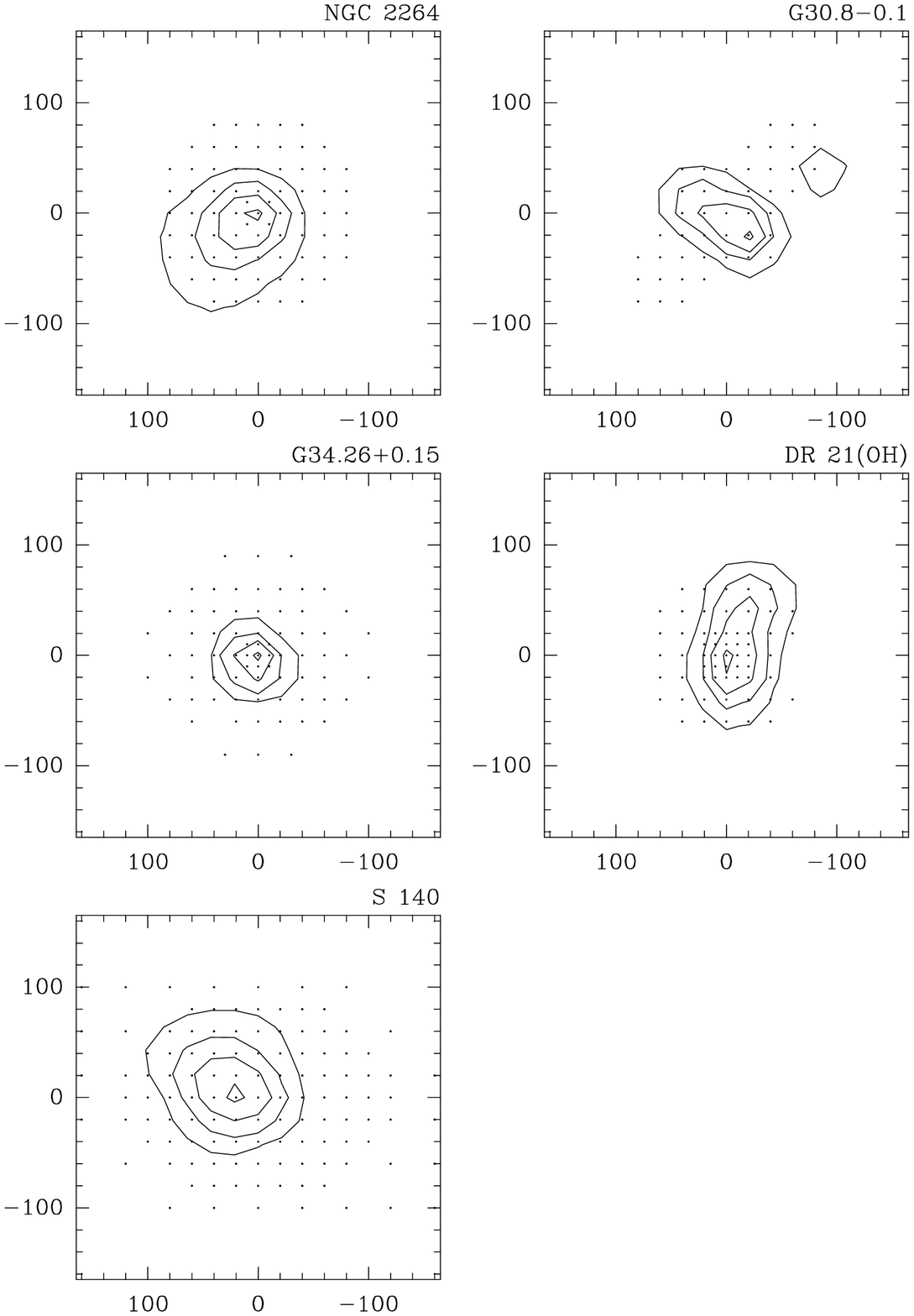}}
\vskip -20mm
\caption{Images of five sources reconstructed by the method of maximum 
entropy. X and Y axes plot, respectively, the RA and DEC offsets in arcsec
from the central positions presented in Table 1. The observed positions are
marked by dots.}
\end{center}
\end{figure*}

The images are shown in Fig. 7.

The resolution of the maps was estimated as follows: at each point of a map,
we subtracted the contribution from the source, convolved with the beam,
leaving only noise. Then we added the contribution from a point source with
flux equal to the observed flux, located at the peak of the reconstructed
image, and convolved with the beam. Finally, we applied the image
reconstruction procedure to this map. The size of the obtained image was
adopted as the map resolution. The resolution of all maps proved to be about
15~--~25 arcsec, being significantly less then the sizes of the sources.
Therefore the angular diameters, presented in Table~4, represent the real
sizes of the sources.

\subsection{COMMENTS ON INDIVIDUAL SOURCES}

{\bf NGC~2264.} The peak of CH$_3$CCH emission coincide with the IR source
IRS1~\citep{sargent} and the continuum sources at 130 and
70~$\mu$m~\citep{Schreyer97}.
The CH$_3$CCH source is slightly extended to the southwest from
the peak, and approximately coincides with the CS, C$^{18}$O, CO, and
CH$_3$OH sources~\citep{Schreyer97}, but in our map there is no second peak approximately
$30''$ southwest from IRS1, which is present in the C$^{18}$O, CS, and
CH$_3$OH maps.

{\bf G34.26+0.15.} The CH$_3$CCH source is essentially circular with
a diameter of $50''$. The peak of the emission coincides with an UC HII
region~\citep{Benson84} and the SO peak~\citep{Carral87}, but it is shifted approximately
$10''$ south-west from the HCO$^+$ peak.

{\bf DR~21(OH).} The methyl acetylene peak coincides with the continuum
sources MM1 and MM2~\citep{Padin89} and the peaks of H$_2$CO, C$^{18}$O,
and CS emission~\citep{Jhonston84,Mangum91}. The methyl acetylene source is extended
in the north--south direction by more than 2 arcmin, like the C$^{18}$O
source~\citep{Padin89}. Hovever, due to the lack of time we could not map the whole
source, and the emission does not vanish at the northern edge of our map.

{\bf S~140.} The source is slightly extended in the northeast--southwest
direction; the peak of the emission coincides with the IR source IRS1~\citep{Evans}.
Our map coincides fairly good  with the $^{12}$CO maps~\citep{Snell84b,Minchin93}.
At the same time,
the peak of CH$_3$CCH emission is located 10--15 arcsec to the east from
the $^{13}$CO, C$^{18}$O, and HCO$^+$ peaks~\citep{Minchin95,Wilner94}.

\subsection{SOURCE PARAMETERS DERIVED FROM THE MAPS}

\begin{table*}
\begin{center}
\caption{Parameters of the mapped sources. Column 1, source; column 2,
distance; column 3, angular diameter; column 4, linear diameter;
column 5, virial mass; column 6, mass, derived from $^{13}$CO observations;
column 7, density, estimated from virial mass; column 8, CH$_3$CCH abundance,
estimated from virial mass and CH$_3$CCH column density.}
\small
\vskip 5pt
\begin{tabular}{|lccccccc|}
\hline
Source         &d     &$\Theta$& D              &M$_{vir}$ &M$_{\rm CO}$ &n$_{\rm H_2}$&CH$_3$CCH\\
               &(kpc) & ('')  &  (cm)        &(M$_{\odot}$)&(M$_{\odot}$)&cm$^{-3}$  &abundance\\
\noalign{\smallskip}
\hline\noalign{\smallskip}
NGC 2264       &0.7   &57     & 6.0$\cdot$10$^{17}$& 89    & 40          &4.8$\cdot10^{5}$&1.6$\cdot10^{-9}$\\
G30.8-0.1$^1$  &6.3   &53     & 5.0$\cdot$10$^{18}$& 6208  & 4919        &5.6$\cdot10^{4}$&4.7$\cdot10^{-9}$\\
G34.26+0.15    &3.8   &42     & 2.4$\cdot$10$^{18}$& 2332  & 528         &1.9$\cdot10^{5}$&3.6$\cdot10^{-9}$\\
DR 21 (OH)$^1$ &3.0   &56     & 2.5$\cdot$10$^{18}$& 1336  & 1941        &9.6$\cdot10^{4}$&7.7$\cdot10^{-9}$\\
S 140          &0.9   &49     & 3.6$\cdot$10$^{17}$& 135   & 67          &5.5$\cdot10^{5}$&2.1$\cdot10^{-9}$\\
\hline\noalign{\smallskip}
\end{tabular}
\end{center}

$^1$--Unreliable parameters (see sect. "The mapping technique").
\end{table*}

We derived the virial masses of the mapped sources assuming that each source
is a uniform optically thin nonmagnetized sphere of a constant density.
According to~\citet{cesaroni}, under these assumptions the virial mass can be calculated
from the formula
\begin{equation}
M_{vir}(M_\odot)=0.509 \cdot d \cdot \Theta_S \cdot \Delta v^2_{1/2}
\end{equation}
where $\Theta_S$ is the source angular diameter in arcsec, $\Delta v^2_{1/2}$
is the full-width at half maximum (FWHM) in km~s$^{-1}$, $d$ is the source
distance in kpc, $M_{vir}(M_\odot)$ is the cloud virial mass in $M_\odot$.
The angular diameter of each source was estimated as follows: first,
we calculated the map area (in arcsec$^2$) within the contour that
corresponds to the half-maximum intensity. This area was multiplied
by $\cos \delta$ to obtain the source area in the sky $A$. The angular
diameter was calculated using the formula $\Theta^2 = 4A/\pi$, and 
additionally multiplied by 1.155 to obtain the sphere diameter from
the vizual diameter~\citep{panagia}. The source distance was derived employing
the Galactic rotation curve from~\citet{brand&bl}.

The masses, determined with Eq. (2), are presented in Table 4. In addition,
we estimated the masses of the same sources using the $^{13}$CO column
densities, assuming that the $^{13}$CO and CH$_3$CCH sources are spatially
coincident and the 1--0 $^{13}$CO lines are optically thin. The latter masses
are also presented in Table 4. For all sources, except G34.26+0.15,
the virial masses coincide with the masses derived from $^{13}$CO column
densities within a factor of two. The comparison of the intensities of
the $^{13}$CO and C$^{18}$O lines in G34.26+0.15, presented in~\citet{little},
leads to the conclusion that the $^{13}$CO lines are not optically thin.
Hence, the derived $^{13}$CO column density and mass of this source are
underestimated.

Since two different methods yield similar mass estimates of four sources,
we believe that these estimates are correct.

Assuming that the masses of the sources are equal to their virial masses
and using the source sizes and methyl acetylene column densities from
Tables 3 and 4, we estimated the gas densities and methyl acetylene
abundances. The estimates are presented in Table 4. The densities appeared
to be close to $10^5$~cm$^{-1}$, and methyl acetylene abundances~--- of
the order of several units x $10^{-9}$. So determined methyl acetylene
abundances within a factor of two coincide with those presented in Table 3
in all sources except G34.26+0.15. We have already mentioned
that the $^{13}$CO column density in this source is probably underestimated.
In this case the methyl acetylene abundance, presented in Table 3,
is overestimated, and the value of $3.6\times 10^{-9}$, presented in Table 4,
is more correct.

Our methyl acetylene abundances are in agreement with the abundances
estimated for warm gas in Orion~\citep{blake,kuiper},
Sgr~B2~\citep{cherch,kuiper}, S~140~\citep{kuiper},
and with the abundance, which can be estimated for the cold cloud TMC-1
from the observations of C$^{18}$O and methyl acetylene~\citep{pratap}.

\subsection{SIMULATIONS OF CHEMICAL EVOLUTION}

\begin{table}
\begin{center}
\caption{Low-metal initial abundances of chemical elements relative to
hydrogen (H$_2$).
\label{elements}
}
\vskip 5mm
\begin{tabular}{|c|c|c|}
\hline\noalign{\smallskip}
Element&C/O=0.41&C/O=0.8\\
\noalign{\smallskip}
\hline\noalign{\smallskip}
 He      &1.4000$\cdot10^{-1}$&1.4000$\cdot10^{-1}$\\
 N       &2.1400$\cdot10^{-5}$&2.1400$\cdot10^{-5}$\\
 O       &1.7600$\cdot10^{-4}$&1.7600$\cdot10^{-4}$\\
 C$^+$   &7.3000$\cdot10^{-5}$&1.4080$\cdot10^{-4}$\\
 S$^+$   &8.0000$\cdot10^{-8}$&8.0000$\cdot10^{-8}$\\
 Si$^+$  &8.0000$\cdot10^{-9}$&8.0000$\cdot10^{-9}$\\
 Fe$^+$  &3.0000$\cdot10^{-9}$&3.0000$\cdot10^{-9}$\\
 Na$^+$  &2.0000$\cdot10^{-9}$&2.0000$\cdot10^{-9}$\\
 Mg$^+$  &7.0000$\cdot10^{-9}$&7.0000$\cdot10^{-9}$\\
 e       &7.3107$\cdot10^{-5}$&7.3107$\cdot10^{-5}$\\
 P$^+$   &3.0000$\cdot10^{-9}$&3.0000$\cdot10^{-9}$\\
 Cl$^+$  &4.0000$\cdot10^{-9}$&4.0000$\cdot10^{-9}$\\
\noalign{\smallskip}
\hline\noalign{\smallskip}
\end{tabular}
\end{center}
\end{table}

Using the derived gas temperatures and densities we modelled the chemical
evolution of the mapped objects. The chemical reaction network corresponded
to the so-called "new standard model"~\citet{lee&bettens}). The code
for the simulations was kindly presented by R. Terzieva from Prof. Herbst's
group in the Ohio University. The model took into consideration only
gas-phase chemistry without any account for dust grains. The initial set of
chemical elements corresponded to the low metal abundance~\citep{lee&bettens}
and is presented in Table 5. The simulations were performed for two C/O ratios,
0.41 and 0.8. The results for NGC~2264 are shown in Fig.~8; the results
for other sources are similar. For $\rm C/O=0.41$ the CH$_3$CCH abundance reaches
its maximum at $t_m \approx 6 \times 10^4$~years since the beginning of
the evolution; later the abundance decreases by several orders of magnitude
and reaches steady-state at $t\ge 10^7$ years. When C/O increases to 0.8,
both $t_m$ and the maximum CH$_3$CCH abundance decreases. The comparison of
the observed CH$_3$CCH abundance (Table 3) with the results of our simulations
shows that it is the abundance that arises around the peak at
$t_m \approx 6\times 10^4$~years is characteristic for molecular clouds. The
same time dependences are typical for many other molecules~\citep{lee&bettens}.
These results implies a large amount of "chemically young" gas in giant
molecular clouds.

\citet{kuiper&langer} in their study of dark clouds
considered the model
of chemically different shells. In this model, the source consists of
a chemically more evolved core, surrounded by a less evolved shell with high
abundances of CH$_3$CCH and other molecules. Such differentiation can be
caused by the accretion of rarefied gas. Chemical processes in rarefied
gas proceed much more slowly than in the core;
in addidion, ultraviolet radiation quickly
destroys the formed molecules. Due to a stable influx of "chemically
young" substance the shell chemical composition is characterized by high
abundances of CH$_3$CCH and other molecules. However, in neither of
the mapped sources could we find any sign of a shell structure. Apparently,
this model is not applicable to warm clouds. Note, however, that in all
sources, except NGC~2264, we could merely not detect the shell structure
because of insufficient spatial resolution.

Another possible explanation of such "chemical" age lie in the fact that
during a time of $\approx 10^5$ years luminous stars may form in the cloud
cores. Radiation of these stars leads to the dissipation of the cores.
One more explanation is the fact that some processes (e.g., eddy
diffusion) may increase the C and C$^+$ abundances in the clouds at late
stages of their evolution, which in turn increases abundances of different
molecules. The review of these and some other possibilities is presented
by \citet{bergin}.

\begin{figure}
\begin{center}
\includegraphics[width=1\linewidth]{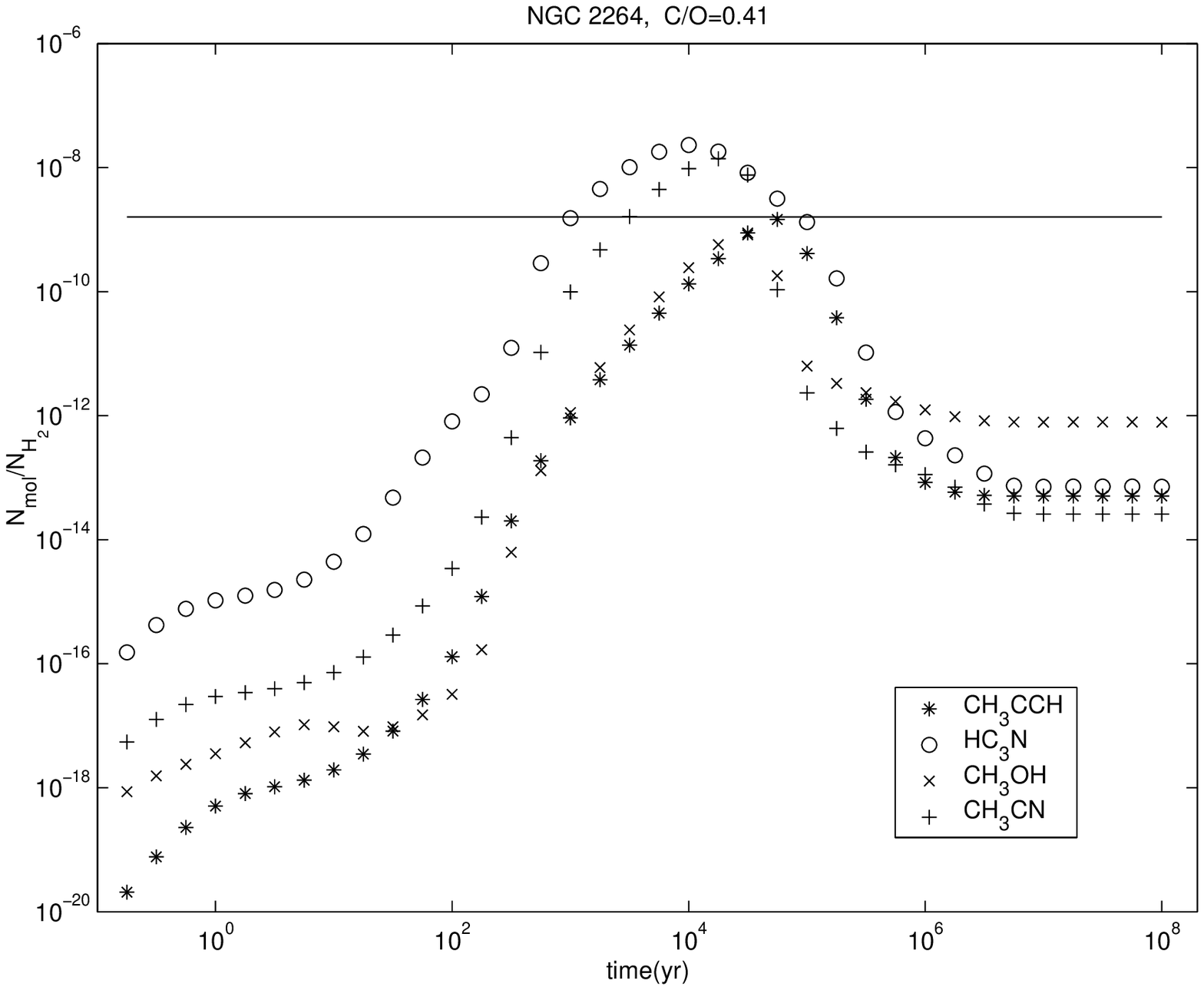}
\includegraphics[width=1\linewidth]{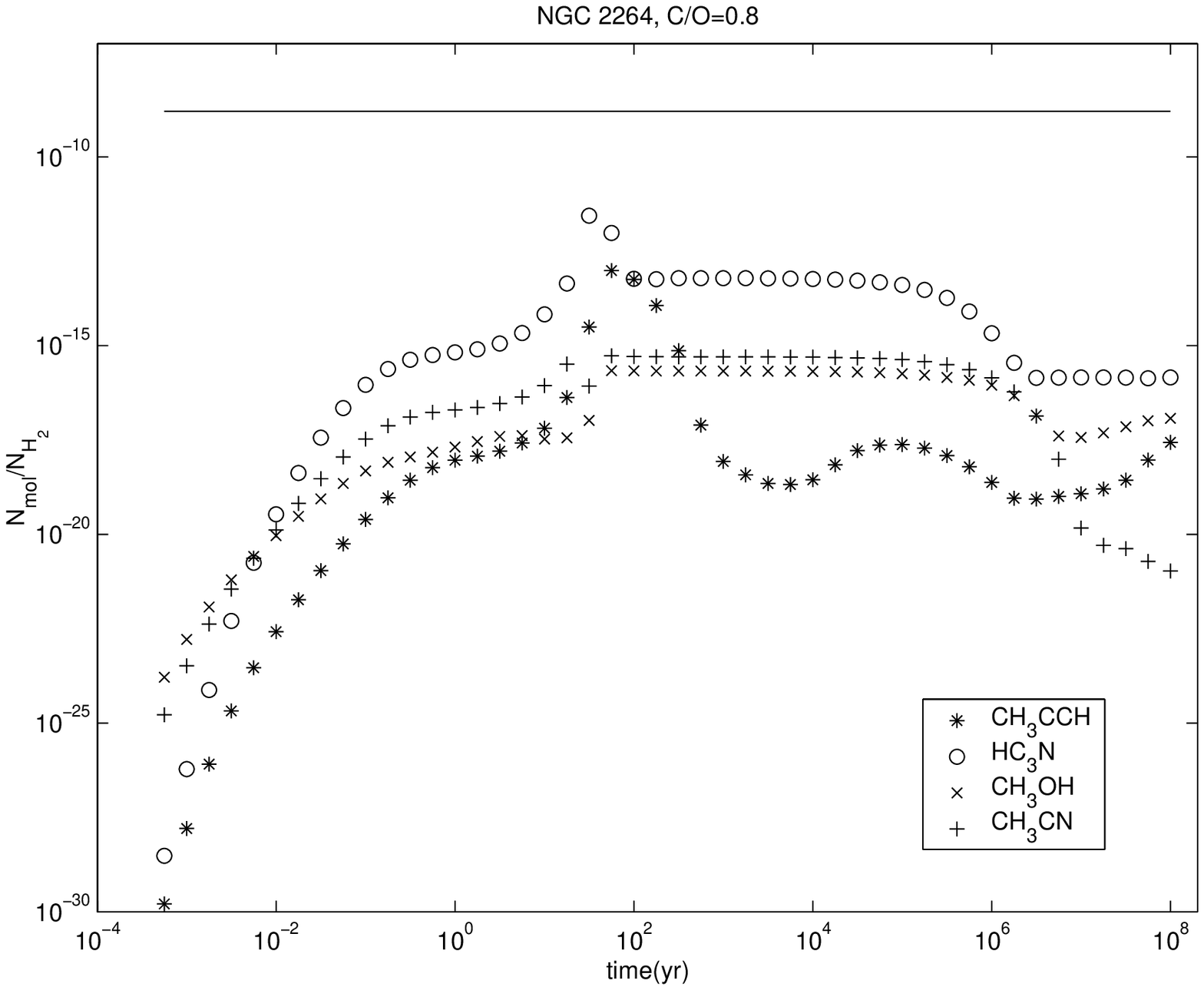}
\caption{The results of chemical simulations for NGC~2264.}
\end{center}
\end{figure}

\section{CONCLUSIONS}

As a result of a survey of Galactic star-forming regions in the methyl
acetylene lines $6_K-5_K$ at 102~GHz and $5_K-4_K$ at 85~GHz, we have detected
44 sources at 102~GHz and 25 sources at 85~GHz.

Using rotational diagrams, we estimated the gas kinetic temperatures
and methyl acetylene abundances. The kinetic temperatures appeared to be
of the order of 20~--~60~K and coincide within errors with the temperatures
that had been estimated from the observations of methanol, methyl acetylene,
and ammonia. In most sources methyl acetylene column densities lie in the 
range $1-5\times 10^{14}$~cm$^{-3}$. The emission of hot cores,
compact massive regions with gas temperatures above 100~K, was not found.

Using the maximum entropy images of five sources, we estimated their sizes,
masses, and densities. The sizes proved to be about 0.1~---~1 pc, the masses~---~hundreds
and thousands $M_\odot$, and the densities~---~about $10^5$~cm$^{-3}$.

Thus, in the lines of methyl acetylene we observed warm and dense clouds.

Simulations of chemical evolution show that the characteristic methyl
acetylene abundances in these clouds corresponds to the gas chemical age
$\approx 6\times 10^4$~years.

\begin{center}ACKNOWLEDGMENTS\end{center}

The authors are grateful to the Onsala Observatory staff for help
during the observations, R. Terzieva, who made available the code for
the simulations of the chemical evolution of molecular clouds,
and Dr. V.I.~Slysh for helpful discussion. The work was partially supported
by the Russian Foundation for Basic Research (grants no. 98-02-16916
and 01-02-16902), project no. 315 "Radio Astronomy Educational and Scientific
Center, and the INTAS grant no. 97-11451. The Onsala Space Observatory
is the Swedish National Facility for Radio Astronomy, and is operated
by the Chalmers University of Technology, G\"oteborg, Sweden, with
financial support from the Swedish Natural Science Research Council
and the Swedish Board for Technical Development.

\end{document}